\documentstyle[pra,eqsecnum,aps]{revtex}
\def\displayfrac#1#2{\frac{\displaystyle #1}{\displaystyle #2}}
\begin{document}
\draft
\title{Characterisations of Classical and Non-classical
         states of Quantised Radiation}
\author{Arvind\cite{email}}
\address{Department of Physics\\
Indian Institute of Science,  Bangalore - 560 012, India}
\author{N. Mukunda\cite{jncasr}}
\address{Center for Theoretical Studies and Department of Physics\\
Indian Institute of Science,  Bangalore - 560 012, India}
\author{R. Simon}
\address{Institute of Mathematical Sciences\\
C. I. T. Campus, Madras - 600 113, India}
\date{\today}
\maketitle
\pacs{03.65.Fd, 42.50.Dv, 42.50.Lc}
\begin{abstract}
A new operator based condition for distinguishing classical
from non-classical states of quantised radiation is
developed.  It exploits the fact that the normal ordering
rule of correspondence to go from classical to quantum
dynamical variables does not in general maintain
positivity. It is shown that the approach naturally leads
to distinguishing several layers of increasing
nonclassicality, with more layers as the number of modes
increases. A generalisation of the notion of subpoissonian
statistics for two-mode radiation fields is achieved by
analysing completely all correlations and fluctuations in
quadratic combinations of mode annihilation and creation
operators conserving the total photon number.  This
generalisation is nontrivial and intrinsically two-mode as
it goes beyond all possible single mode projections of the
two-mode field. The nonclassicality of pair coherent
states, squeezed vacuum and squeezed thermal states is
analysed and contrasted with one another, comparing the
generalised subpoissonian statistics with extant signatures
of nonclassical behaviour.
\end{abstract}
\section{Introduction}
\setcounter{equation}{0}
Electromagnetic radiation is intrinsically quantum mechanical in
nature. Nevertheless it has been found extremely fruitful, at
both conceptual and practical levels, to designate certain
states of quantised radiation as being essentially
``classical'', and others as being
``non-classical''~\cite{walls-nature}.  It is the
latter that show the specific quantum features of radiation most
sharply.  Some of the well known signs of nonclassicality in
this context are quadrature squeezing~\cite{squeezing},
antibunching~\cite{antibunching} and
subpoissonian photon statistics~\cite{sub-poissonian}.

The purposes of this paper are to present a new physically
equivalent way of distinguishing classical from nonclassical
states of radiation, dual to the customary definition and based
on operator properties; to point out the existence of several
levels of classical behaviour, with a structure that gets
progressively more elaborate as the number of modes increases;
and finally to give a complete discussion of signatures of
nonclassical photon statistics for two-mode fields, working at
the level of fluctuations in photon numbers.

The contents of this paper are arranged as follows In section II
we develop a criterion based on operator expectation values, to
distinguish between classical and nonclassical states of
radiation. The basic idea is that the normal ordering rule of
correspondence between classical dynamical variables and quantum
operators, while being linear and translating reality into
hermiticity, does not respect positivity. If this potential
nonpositivity does not show up in the expectation values of
operators in a certain state, then that state is classical;
otherwise it is nonclassical.
Section III explores this new approach
further and shows that, as the number of independent modes
increases, the classification of quantum states gets
progressively finer; several levels of nonclassicality emerge.
This is shown in detail for one and two mode fields, and then
the trend becomes clear. Section IV analyses in complete detail
the properties of two mode photon number fluctuations,
stressing the freedom to choose any normalised linear
combination of the originally given modes as a variable single
mode. The well known Mandel parameter criterion~\cite{mandel-Q}
 for sub-Poissonian
statistics for a single mode field is extended in full
generality to a matrix inequality in the two-mode case. It is
shown that certain consequences of this inequality transcend the
set of all single mode projections of it, and are thus
intrinsically two-mode in character. Explicit physically
interesting examples of this situation are provided, and the well
known pair coherent states are also examined from this point of
view. Section V presents some concluding remarks.
\section{The distinction between classical and
non-classical states - an operator criterion}
\setcounter{equation}{0}

We deal for simplicity with states of a single mode radiation
field, though our arguments generalise immediately to any number
of modes.  The photon creation and annihilation operators
$\hat{a}^{\dag}$ and $\hat{a}$ obey the customary commutation
relation
\begin{eqnarray}
\mbox{}[\hat{a},\;\hat{a}^{\dag}] \equiv \hat{a}\;\hat{a}^{\dag} -
\hat{a}^{\dag}\;\hat{a}=1\cdot
\end{eqnarray}
\noindent
The coherent states $|z\rangle$ are right eigenstates of
$\hat{a}$ with (a generally complex) eigenvalue $z$; they are
related to the states $|n\rangle$ of definite photon number
(eigenstates of $\hat{a}^{\dag}\;\hat{a}$) in the standard way:
\begin{eqnarray} |z\rangle &=& e^{-|z|^2/2}
\sum\limits^{\infty}_{n=0}
\displayfrac{z^n}{\sqrt{n!}} |n\rangle\;,\nonumber\\
\hat{a}|z\rangle &=& z|z\rangle\;,\nonumber\\
\langle z^{\prime}|z\rangle &=&
\exp\left(-\frac{1}{2}|z^{\prime}|^2-\frac{1}{2}|z|^2 + z^{\prime *}z
\right)\;;\nonumber\\ |n\rangle &=&
\displayfrac{\left(\hat{a}^{\dag}\right)^n}{\sqrt{n!}}|0
\rangle\;,\nonumber\\ \hat{a}^{\dag}\hat{a}|n\rangle &=&
n|n\rangle\;,\nonumber\\ \langle n^{\prime}|n\rangle &=&
\delta_{n^{\prime}n}\cdot
\end{eqnarray}
\noindent
A general (pure or mixed) state of the one-mode field is
described by a corresponding normalised density matrix
$\hat{\rho}$:
\begin{eqnarray}
\hat{\rho}^{\dag} =
\hat{\rho} \geq 0,\quad\mbox{Tr}\;\hat{\rho} = 1\cdot
\end{eqnarray}
\noindent
It can be expanded in the so-called diagonal coherent state
representation~\cite{klouder-sudershan-glouber}:
 \begin{eqnarray} \hat{\rho}
&=&\int\;\displayfrac{d^2 z}{\pi}\;\phi(z)\; |z\rangle\langle
z|\;,\nonumber\\ &&\int\;\displayfrac{d^2 z}{\pi}\;\phi(z) =
1\cdot
\label{diag-coh-def}
 \end{eqnarray}
\noindent
While hermiticity of $\hat{\rho}$ corresponds to reality of the
weight function $\phi(z)$, the latter is in general a singular
mathematical quantity, namely a distribution of a well-defined
class.

The conventional designation of $\hat{\rho}$ as being classical or
nonclassical is based on the properties of $\phi(z)$.  Namely,
$\hat{\rho}$ is said to be classical if $\phi(z)$ is everywhere
non-negative and not more singular than a delta
function~\cite{walls-nature}:
\begin{eqnarray}
\hat{\rho}\;\mbox{classical}&\Leftrightarrow & \phi(z)\geq 0,\;
\mbox{no worse than delta function}\;,\nonumber\\
\hat{\rho}\;\mbox{nonclassical}&\Leftrightarrow & \phi(z)
\not\geq 0\cdot
\label{diag-coh-nonc-cond}
\end{eqnarray}
\noindent
It is clear that the conditions to be classical involve an
infinite number of independent inequalities, since $\phi(z)\geq
0$ has to be obeyed at each point $z$ in the complex plane.
This is true despite the fact that the condition $\hat{\rho}
\geq
0$ means that the ``values'' of $\phi(z)$ at different points
$z$ are not quite ``independent''.  One realises this by
recalling that every classical probability distribution over the
complex plane is certainly a possible choice for $\phi(z)$, with
the corresponding $\hat{\rho}$ being classical.

The above familiar definition of classical states deals directly
with $\hat{\rho}$ and $\phi(z)$.  Now we develop a dual, but
equivalent, definition based on operators and their expectation
values. As is well known, the
representation~(\ref{diag-coh-def}) for
$\hat{\rho}$ is closely allied to the normal ordering rule for
passing from classical c-number dynamical variables to quantum
operators.  Within quantum mechanics we know that an operator
$\widehat{F}$ is completely and uniquely determined by its
diagonal coherent state matrix elements (expectation values)
$\langle z|\widehat{F}|z\rangle$.  Moreover, hermiticity of
$\widehat{F}$ and reality of $\langle z|\widehat{F}|z\rangle$
are precisely equivalent.  Any (real) classical function
$f(z^*,z)$ determines uniquely, by the normal ordering rule of
placing $\hat{a}^{\dag}$ always to the left of $\hat{a}$ after
substituting $z\rightarrow \hat{a}$ and $z^*\rightarrow
\hat{a}^{\dag}$, a corresponding (hermitian) operator
$\widehat{F}_N$ as follows:

\noindent
\underline{Normal ordering rule}
 \begin{eqnarray} f(z^*, z)&\longrightarrow &
\widehat{F}_N\;,\nonumber\\
\langle z|\widehat{F}_N|z\rangle &=& f(z^*,z)\;,\nonumber\\
f\;\mbox{real}&\Leftrightarrow & \widehat{F}_N\;\mbox{hermitian}
\label{norm-order}
\end{eqnarray}
\noindent
The connection with the representation~(\ref{diag-coh-def})
for $\hat{\rho}$ is
given by
 \begin{eqnarray}
\mbox{Tr}\;\left(\hat{\rho}\;\widehat{F}_N\right) =
\int\;\displayfrac{d^2z}{\pi}\;\phi(z)\;f(z^*,z)\cdot
\label{expect-diagcoh}
\end{eqnarray}
It is an important property of the normal ordering rule that,
while it translates classical reality to quantum hermiticity,
{\em it does not preserve positive semidefiniteness}.
More explicitly, while by eq.~(\ref{norm-order})
 $\widehat{F}_N\geq 0$ implies
$f(z^*,z)\geq 0$, {\em the converse is not true}.  Here are some
simple examples of nonnegative classical real $f(z^*,z)$ leading
to indefinite hermitian $\widehat{F}_N$:
\begin{eqnarray}
f(z^*,z) &=& (z^* +z)^2\longrightarrow \widehat{F}_N =
\left(\hat{a}^{\dag} + \hat{a}\right)^2 - 1\;;\nonumber\\
f(z^*,z) &=& (z^* +z)^4 \longrightarrow \widehat{F}_N =
\left(\left(\hat{a}^{\dag} +\hat{a}\right)^2 -3\right)^2 -
6\;;\nonumber\\ f(z^*,z) &=&
e^{-z^*z}\;\sum\limits^{\infty}_{n=0}\;
\displayfrac{C_n}{n!}\;z^{*n}z^n\rightarrow \widehat{F}_N =
\sum\limits^{\infty}_{n=0}\;C_n|n\rangle\langle n|\cdot
\label{cl-positive-op-indef}
\end{eqnarray}

\noindent
In the last example, the real constants $C_n$ can certainly be
chosen so that some of them are negative while maintaining
$f(z^*,z)\geq 0$; this results in $\widehat{F}_N$ being
indefinite.

We thus see that when the normal ordering rule is used, every
$\widehat{F}_N\geq 0$ arises from a unique $f(z^*,z)\geq 0$, but
some (real) $f(z^*,z)\geq 0$ lead to (hermitian) indefinite
$\widehat{F}_N$.  So in a given quantum state $\hat{\rho}$, the
operator $\widehat{F}_N$ corresponding to a nonnegative real
classical $f(z^*,z)$ could well have a negative expectation
value. {\em If this never happens, then $\hat{\rho}$ is
classical}.  That is, as we see upon combining
eqns.(~\ref{diag-coh-nonc-cond})~(\ref{expect-diagcoh}): if
for every $f(z^*,z)\geq 0$ the corresponding $\widehat{F}_N$ has
a nonnegative expectation value even though $\widehat{F}_N$ may
be indefinite, then $\hat{\rho}$ is classical.  Conversely,
$\hat{\rho}$ is nonclassical if there is at least one
$f(z^*,z)\geq 0$ which leads to an indefinite $\widehat{F}_N$
whose expectation value is negative.

We can convey the content of this dual operator way of defining
classical states also as follows: while the normal ordering rule
allows for the appearance of ``negativity'' in an operator
$\widehat{F}_N$ even when none is present in the corresponding
classical $f(z^*,z)$, in a classical state such negativity never
shows up in expectation values.

Purely by way of contrast, we compare the above with what
obtains when the antinormal ordering rule - substituting
$z\rightarrow\hat{a},\;z^*\rightarrow \hat{a}^{\dag}$ followed
by placing $\hat{a}$ to the left, $\hat{a}^{\dag}$ to the right
- is used to pass from classical $f(z^*,z)$ to quantum
$\widehat{F}$~\cite{mehata-sudarshan}.  In place of
eqns.(~\ref{norm-order})~(\ref{expect-diagcoh}) we have:

\noindent
\underline{Antinormal ordering rule}
 \begin{eqnarray}
\mbox{real}\;f(z^*,z)&\rightarrow &\mbox{hermitian}\;\widehat{F}_A
= \int\;\displayfrac{d^2z}{\pi}\;f(z^*,z)\;|z\rangle\langle
z|\;, \nonumber\\
\mbox{Tr}\;\left(\hat{\rho}\;\widehat{F}_A\right) &=&
\int\;\displayfrac{d^2z}{\pi}\;\langle z|\hat{\rho}|z\rangle\;
f(z^*,z)\cdot \end{eqnarray}
\noindent
Now the situation is that $f(z^*,z)\geq 0$ certainly implies
$\widehat{F}_A\geq 0$, but some $\widehat{F}_A\geq 0$ arise from
indefinite classical $f(z^*,z)$.  A simple example is:
 \begin{eqnarray} f(z^*,z) = (z^*
+z)^2-1\longrightarrow
\widehat{F}_A =\left(\hat{a}^{\dag}
+ \hat{a}\right)^2\cdot \end{eqnarray}
\noindent
Thus the function and operator relationships are opposite to
what we found in the normal ordering case.  The direct
characterization of classical states (whose definition in any
case is a convention based on the normal ordering rule) is
however not very convenient with the antinormal ordering
convention.

In passing we may note an interesting aspect of the Weyl or
symmetric rule of ordering~\cite{weyl-book}, and the
associated Wigner
distribution description of a state
$\hat{\rho}$~\cite{wigner-1932}.  Here it is
more convenient to deal with the real and imaginary parts of $z$
and $\hat{a}$:
 \begin{eqnarray} z &=&
\frac{1}{\sqrt{2}}\;(q+i\;p),\quad z^* =
\frac{1}{\sqrt{2}}(q-i\;p)\;;
\nonumber\\ \hat{a} &=&\frac{1}{\sqrt{2}}\;(\hat{q}
+i\;\hat{p}),\quad \hat{a}^{\dag}
=\frac{1}{\sqrt{2}}(\hat{q}-i\;\hat{p})\;,\nonumber\\
&&[\hat{q},\;\hat{p}] = i\;\cdot \end{eqnarray} The Weyl
ordering rule maps single classical exponentials into single
operator exponentials,
 \begin{eqnarray}
\exp(i\;\lambda\;q + i\;\mu\;p)\rightarrow
\exp(i\;\lambda\;\hat{q} + i\;\mu\;\hat{p})\;,\quad
\lambda\;\mbox{and}\;\mu\;\mbox{real}\;;
\end{eqnarray}
\noindent
and then extends this by linearity and Fourier transformation to
general functions:

\noindent
\underline{Weyl rule}
 \begin{eqnarray} f(q,p) &=&
\int\limits^{\infty}_{-\infty}d\lambda \int\limits^{\infty}
_{-\infty} d\mu\; \tilde{f}(\lambda,\mu)\exp(i\;\lambda\;q +
i\;\mu\;p) \longrightarrow\nonumber\\ \widehat{F}_W &=&
\int\limits^{\infty}_{-\infty} d\lambda\int\limits
^{\infty}_{-\infty} d\mu \;\tilde{f}(\lambda,\mu)
\exp(i\;\lambda\;\hat{q} +i\;\mu\;\hat{p})\;,\nonumber\\
f\;\mbox{real}&\Longleftrightarrow
&\widehat{F}_W\;\mbox{hermitian}
\label{weyl-rule}
\end{eqnarray}
\noindent
The Wigner distribution $W(q,p)$ for a state $\hat{\rho}$ is
given in terms of the configuration space matrix elements of
$\hat{\rho}$:
 \begin{eqnarray}
 W(q,p) =
\int\limits^{\infty}_{-\infty} dq^{\prime}
\langle q-\frac{1}{2} q^{\prime}|\hat{\rho}|q+\frac{1}{2}
q^{\prime}\rangle\; e^{ipq^{\prime}}\;,
\label{wigner-function}
\end{eqnarray}
\noindent
and the rule for expectation values ties together
eqns.~(\ref{weyl-rule})~(\ref{wigner-function}):
 \begin{eqnarray}
\mbox{Tr}\;\left(\hat{\rho}\;\widehat{F}_W\right) =
\int\limits^{\infty}_{-\infty}\;
\int\limits^{\infty}_{-\infty}\;dq\;dp\;W(q,p)\;f(q,p)\cdot
\end{eqnarray}
\noindent
{}From the algebraic point of view the Weyl rule stands ``midway''
or symmetrically between the normal and the antinormal ordering
rules.  This may lead one to hope that it removes the mismatch
and resolves the problem of preserving positivity in both
directions in passing between $f(q,p)$ and $\widehat{F}_W$.
However this does not happen at all.  As is well known there are
nonnegative $f(q,p)$ yielding indefinite $\widehat{F}_W$, and
nonnegative $\widehat{F}_W$ leading back to indefinite $f(q,p)$.
Here are simple examples:
 \begin{eqnarray}
f(q,p) &=&\delta(q)\delta(p)\longrightarrow \widehat{F}_W =
\mbox{parity operator with eigenvalues}\;\pm 1\;;
\nonumber\\ \widehat{F}_W &=& |1\rangle\langle 1|\longrightarrow
f(q,p) =\frac{2}{\pi} \left(q^2 + p^2 -\frac{1}{2}\right)
\exp\left(-q^2 -p^2\right)\cdot
\end{eqnarray}

While these remarks illuminate in terms of
operator properties the relations among the three ordering
rules, the classification of states $\hat{\rho}$ into classical
and nonclassical ones is based most simply on the normal
ordering rule. It is clear that all these considerations
extend easily to any number of modes of radiation.

While our operator based approach to the identification of
nonclassical states is conceptually complete, we note that
most of the extant criteria of nonclassicality involve a simple
extension of our formalism. Namely, one often has to consider
nonlinear functions of expectation values of several operators,
which cannot be simply expressed as the expectation value of a
single state-independent or autonomous operator. This is always
so when one deals with fluctuations. Thus quadrature squeezed
non-classical states are defined via inequalities involving the
fluctuations $(\Delta \hat{q})^2\/$ and $(\Delta \hat{p})^2\/$.
Similarly amplitude squeezing or sub-Poissonian statistics deals
with the fluctuation in photon number. These remarks apply also
to other criteria of non-classicality such as higher order
squeezing~\cite{hong-mendal}
and the one based on matrices constructed out of
factorial moments of the photon number
distribution~\cite{tara-agarwal}. The one
interesting exception to these  remarks and which is fully
covered by our formalism, is the case of
antibunching~\cite{antibunching}. Here one
is concerned with the expectation values of the single
time-dependent operator
\begin{equation}
\hat{F}(t,t+\tau)=:\hat{I}(t+\tau) \hat{I}(t):-
  :\hat{I}(t) \hat{I}(t):
\end{equation}
where $\hat{I}(t)\/$ is the intensity operator at time $t\/$. In
this sense, the criterion for antibunching is qualitatively
different from the other familiar ones.

\section{Levels of classicality}
\setcounter{equation}{0}
\subsection{The single mode case}
We begin again with the single mode situation and hereafter deal
exclusively with the normal ordering prescription.  (Therefore
the subscript $N$ on $\widehat{F}_N$ will be omitted).  Suppose
we limit ourselves to classical functions $f(z^*,z)$ which are
real, nonnegative and phase invariant, that is, invariant under
$z\rightarrow e^{i \alpha}z$.  An independent and complete set
of these can be taken to be
 \begin{eqnarray}
f_n(z^*,z) = e^{-z^*z}\;z^{*n} z^n/n!\;,\quad
n=0,\;1,\;2,\;\ldots,
\label{cl-fn-examp}
\end{eqnarray}
\noindent
since they map conveniently to the number state projection
operators:
 \begin{eqnarray}
f_n(z^*,z)\longrightarrow \widehat{F}^{(n)} = |n\rangle\langle
n|\;, \quad n=0,\;1,2,\;\ldots
\label{cl-fn-examp-op}
\end{eqnarray}
\noindent
A general real linear combination
$f(z^*,z)=\sum\limits_n\;C_n\;f_n(z^*,z)$, even if nonnegative,
may lead to an indefinite $\widehat{F}$, as seen at
eqn.~(\ref{cl-positive-op-indef}).

If we are interested only in the expectation values of such
variables, we are concerned only with the probabilities $p(n)$
for finding various numbers of photons; for this purpose an
angular average of $\phi(z)$ is all that is required:
 \begin{eqnarray} p(n) &=& \langle
n|\hat{\rho}|n\rangle =\mbox{Tr}\;(\hat{\rho}|n\rangle \langle
n|)\nonumber\\ &=&\int\;\displayfrac{d^2
z}{\pi}\;\phi(z)\;e^{-z^*z} z^{*n} z^n\big/n!  \nonumber\\
&=&\int\limits^{\infty}_0\;dI\;P(I)\;e^{-I}\;I^n/n!\;,\nonumber\\
P(I)&=& \displayfrac{1}{2\pi}\;\int\limits^{2\pi}_0\;d\theta\;
\phi\left(I^{1/2}\;e^{i\theta}\right)\cdot
\label{prob-I}
\end{eqnarray}
\noindent
Now while $\phi(z)\geq 0$ certainly implies $P(I)\geq 0$, the
converse is not true.  Thus one is led to a three-fold
classification of quantum states $\hat{\rho}$~\cite{mehata-1994}:
\begin{eqnarray}
\hat{\rho}\;\mbox{classical}&\Longleftrightarrow & \phi(z)\geq
0\;,\;\mbox{hence}\;P(I)\geq 0\;;\nonumber\\
\hat{\rho}\;\mbox{semiclassical} &\Longleftrightarrow &
P(I)\geq 0\;\mbox{but}\;\phi(z)\not\geq 0\;;\nonumber\\
\hat{\rho}\;\mbox{strongly nonclassical} &\Longleftrightarrow &
P(I)\not\geq 0\;,\;\mbox{so necessarily}\; \phi(z)\not\geq
0\cdot
\label{cl-semicl-noncl}
\end{eqnarray}
\noindent
The previous definition~(\ref{diag-coh-nonc-cond})
 of nonclassical $\hat{\rho}$ based
on $\phi(z)$ alone is now refined to yield two subsets of
states, the semiclassical and the strongly nonclassical.  The
semiclassical states do have the following property:
 \begin{eqnarray}
\hat{\rho}\;\mbox{semiclassical}\Longrightarrow \;\mbox{Tr}
\left(\hat{\rho}\widehat{F}\right)\geq 0 \;\mbox{if}\; f(z^*,z)
=\sum\limits^{\infty}_{n=0}\;C_n\;f_n(z^*,z)\geq 0\cdot
\end{eqnarray}
\noindent
However, in addition, there would definitely be some phase
noninvariant $f(z^*,z)\geq 0$ for which $\widehat{F}$ is
indefinite and $\mbox{Tr}\left(\hat{\rho}\widehat{F}\right) \,<
\,0$.
It is just that this extent of nonclassicality in $\hat{\rho}$
is not revealed by the expectation values
of phase invariant variables, or at the level of the
probabilities $p(n)$~\cite{klouder-sudarshan}.

It is clear that the classification~(\ref{cl-semicl-noncl})
is $U(1)\/$ or phase
invariant. That is, $\hat{\rho}\/$ retains its classical,
semiclassical or strongly non-classical character under the
transformation $ \phi(z) \rightarrow \phi^{\prime}(z)= \phi(z e^{i
\alpha})\/$

As examples of interesting inequalities obeyed if $\hat{\rho}$
is either classical or semiclassical, we may quote the following
involving the factorial moments of the photon number
probabilities $p(n)$:
\begin{eqnarray}
\gamma_m &=&\mbox{Tr}\left(\hat{\rho}
\hat{a}^{\dag m}\hat{a}^m\right) \nonumber  \\
&=&\int\;\displayfrac{d^2z}{\pi}\;\phi(z)\;(z^*z)^m \nonumber \\
&=&\int\limits^{\infty}_0\;d I\;P(I)\;I^m \nonumber \\
&=&\sum\limits^{\infty}_{n=m}\;p(n)\;n!/(n-m)!\;\geq 0,\quad
m=0,\;1,\;2,\ldots;\nonumber \\
\hat{\rho}\;\mbox{classical or semiclassical}&\Longleftrightarrow&
P(I)\geq 0\Longrightarrow
\gamma_m\;\gamma_n\leq\gamma_{m+n} \leq\sqrt{\gamma_{2m}\gamma_{2n}}
\label{two-mode-defns-I}
\end{eqnarray}
\noindent
Violation of any one of these inequalities implies $\hat{\rho}$
is strongly nonclassical.

The inequalities quoted in eq.~(\ref{two-mode-defns-I})
above clearly involve an
infinite subset of the photon number probabilities $p(n)$.
However one can easily construct far simpler inequalities
involving a small number of the $p(n)$'s, violation of any of
which also implies that $\hat{\rho}\/$ is strongly
non-classical. For example,
from eqs.~(\ref{cl-fn-examp})~(\ref{cl-fn-examp-op}), for any
nonnegative integer $n_0\/$ and any real $a, b\/$ we have the
correspondence
\begin{eqnarray}
f(z^{\star},z)&=&
e^{-z^{\star}z}\,\frac{\displaystyle{(z^{\star}z)^{n_0}}}{n_0!}\,
(a+bzz^{\star})^2 \rightarrow \nonumber \\
\hat{F}&=&a^2 \vert n_0\rangle \langle n_0 \vert + 2(n_0+1)ab
\vert n_0+1\rangle \langle n_0 +1 \vert+(n_0+1)(n_0+2)b^2
\vert n_0+2\rangle \langle n_0+2 \vert
\end{eqnarray}
Here $f(z^{\star},z)\/$ is nonnegative while $\hat{F}\/$ is
indefinite if $ab<0$. We then have the result :
\begin{eqnarray}
\hat{\rho} \quad \mbox{classical or semiclassical}\,
\Leftrightarrow P(I)\geq 0 \Rightarrow \nonumber \\
a^2\/ p\/(n_0)+ 2\/(n_0+1) a\/b\/p\/(n_0+1) +
(n_0+1)(n_0+2)\/b^2\/p\/(n_0+2)
\nonumber \\
=\frac{1}{n_0!}
\int\limits_0^{\infty} dI P(I) e^{-I} I^{n_0} (a+bI)^2
 \geq 0
\end{eqnarray}
So again, violation of any of these "local" inequalities in
$p(n)\/$ implies that $\hat{\rho}\/$ is strongly nonclassical.

A physically illuminating example of the distinction between
classical and semiclassical $\hat{\rho}\/$, and passage from
one to the other, is provided by the case of the Kerr
medium. The argument is intricate and rests on two well
known results. The first is Hudson's
theorem~\cite{hudson-th}~: if a (purestate)
wavefunction $\psi_0(q)\/$ has a nonnegative Wigner function
$W_0(q,p)$, then $\psi_0(q)\/$ is Gaussian and conversely;
in that case $W_0(q,p)\/$ is also Gaussian. The second
result is the general connection between $\phi(z)\/$ and
$W(q,p)\/$ for any $\hat{\rho}\/$:
\begin{equation}
W(q,p)=2 \int \frac{d^2 z^{\prime}}{\pi}
e^{-2\vert z-z^{\prime}\vert^2} \phi(z^{\prime}),
 \quad z=\frac{1}{\sqrt{2}}(q+ip).
\label{kerr-wigner}
\end{equation}
This means that for classical $\hat{\rho} \/$ with $\phi(z)
\geq 0\/$,
$W(q,p) \geq 0 \/$ as well. Now imagine a single mode
radiation field in an initial coherent state $\vert z_0
\rangle \/$ with $\phi_0(z)
= \pi \delta^{(2)} (z-z_0)\/$, incident upon a
Kerr medium~\cite{kerr-medium}.
This initial state is pure, classical,and has a Gaussian
wave function $\psi_0(q)\/$. The Kerr medium Hamiltonian is
of the form
\begin{equation}
H_{\mbox{Kerr}}= \alpha\,\hat{a}^{\dagger}\hat{a} +
\beta(\hat{a}^{\dagger}\hat{a})^2
\end{equation}
Clearly the number states $\vert n \rangle\/$ are eigenstates
of this Hamiltonian. Therefore the Poissonian photon
number distribution
\begin{equation}
p(n) = e^{-I_0} I_0^n /n!, \quad I_0=z_0^{\star}z_0,
\end{equation}
of the input state $\vert z_0 \rangle \/$ is preserved
under passage through the Kerr medium. Likewise the
funtion $P(I)=\delta(I-I_0)\/$ is left unaltered. Therefore
the output state $\vert \psi \rangle \/$, which of course
is pure, is either classical or semiclassical. However the
form of $ H_{\mbox{Kerr}}\/$ shows that the output
wavefunction is non Gaussian. Therefore by Hudson's theorem
the corresponding $W(q,p)\/$ must become negative
somewhere. Therefore by eq.~(\ref{kerr-wigner})
the output $\phi(z)\/$
cannot be nonnegative. Thus passage through the Kerr medium
converts an incident coherent state, which is classical,
into a final state which is semiclassical.
\subsection{The Two-mode case}
Now we sketch the extension of these ideas to the two-mode case.
Here the operator commutation relations, number and coherent
states, and the diagonal representation for $\hat{\rho}$, are as
follows:
 \begin{eqnarray}
[\hat{a}_r,\hat{a}_s^{\dag}] &=&
\delta_{rs},\;[\hat{a}_r,\hat{a}_s] =
[\hat{a}_r^{\dag},\hat{a}^{\dag}_s]=0,\quad r,\;s=1,\;2\;;\nonumber \\
|n_1,n_2\rangle&=&\left(n_1!n_2!\right)^{-1/2}\left(\hat{a}_1^{\dag}
\right)^{n_1}\left(\hat{a}_2^{\dag}
\right)^{n_2}|0,\;0\rangle\;, \nonumber \\
\left(\hat{a}^{\dag}_1\hat{a}_1\;\mbox{or}\;\hat{a}^{\dag}_2\hat{a}_2
\right) |n_1,n_2\rangle &=& \left(n_1\;\mbox{or} \;n_2\right)
|n_1,n_2\rangle\;; \nonumber \\ |\underline{z}\rangle &=&
|z_1,z_2\rangle = \exp
\left(-\frac{1}{2}\underline{z}^{\dag}\underline{z}\right)
\sum\limits^{\infty}_{n_{1},n_{2}=0}
\displayfrac{z_{1}^{n_{1}}z_2^{n_{2}}}
{\left(n_1! n_2!\right)^{1/2}} |n_1, n_2\rangle\;;\nonumber\\
\hat{\rho} &=& \int\;d\mu(\underline{z})\;\phi(\underline{z})
|\underline{z}\rangle\langle\underline{z}|\;,\nonumber\\
d\mu(\underline{z}) &=& d^2z_1\;d^2z_2\big/\pi^2\cdot
\end{eqnarray}
\noindent
It is convenient at this point to go when necessary beyond
purely real classical functions $f(\underline{z})$ in applying
the normal ordering rule to obtain corresponding operators.
{}From the general number states matrix elements of $\hat{\rho}$
we read off some operator correspondences generalising
eqns.~(\ref{cl-fn-examp})~(\ref{cl-fn-examp-op}):
 \begin{eqnarray} \langle n_3,
n_4|\hat{\rho}|n_1, n_2 \rangle &=&
\mbox{Tr}(\hat{\rho}|n_1,n_2\rangle\langle n_3,n_4|)\nonumber\\
&=&\int\;d\mu(\underline{z})\,\phi(\underline{z})\;
e^{-\underline{z}^{\dag}\underline{z}}
\displayfrac{z^{*n_{1}}_1z^{*n_{2}}_2z_1^{n_{3}}z_2^{n_{4}}}
{\sqrt{n_1! n_2! n_3! n_4!}}\Longrightarrow\nonumber\\
e^{-\underline{z}^{\dag}\underline{z}}
\displayfrac{z_1^{*n_{1}}z_2^{*n_{2}}z_1^{n_{3}}z_2^{n_{4}}}
{\sqrt{n_1! n_2! n_3! n_4!}}&\longrightarrow&|n_1, n_2\rangle
\langle n_3, n_4|\cdot
\label{examp-rho-matelements}
\end{eqnarray}

For one mode the phase transformations form the group $U(1)$.
For two modes this generalises to the group $U(2)$ of (passive)
transformations mixing the two orthonormal single photon modes.
At the operator level this means that the annihilation operators
$\hat{a}_r$ experience a general $U(2)$ matrix transformation
conserving total photon number.  Moreover these transformations
are unitarily implemented on the two-mode Hilbert
space~\cite{pramana-review}:
 \begin{eqnarray} u
=\left(u_{rs}\right)\;\epsilon\;U(2)\;:&&\nonumber\\ &{\cal
U}(u)\;\hat{a}_r\;{\cal U}(u)^{-1} =& u_{sr}\;\hat{a}_s\;,
\nonumber\\
&{\cal U}(u)\;\hat{a}_r^{\dag}\;{\cal U}(u)^{-1} =&
u^*_{sr}\;\hat{a}^{\dag}_s\;,\nonumber\\ &{\cal U}(u)\;{\cal
U}(u)^{\dag} =& 1\;,\nonumber\\ &{\cal U}(u^{\prime})\;{\cal
U}(u) =& {\cal U}(u^{\prime}u)\cdot \end{eqnarray}
\noindent
For later use we give here the actions of these unitary
operators ${\cal U}(u)$ on monomials formed out of
$\hat{a}^{\dag}_r$ and $\hat{a}_r$, on the number states and on
coherent states:
 \begin{eqnarray}
u=e^{i\alpha}
a\;\epsilon\;U(2),\;a\;\epsilon\;SU(2)\;&:&\nonumber\\ {\cal
U}(u)\displayfrac{\hat{a}_1^{\dag j+m}\hat{a}_2^{\dag j-m}}
{\sqrt{(j+m)!(j-m)!}} {\cal U}(u)^{-1} &=& e^{-2i\alpha
j}\sum\limits_{m^{\prime}} D^{(j)}_{m^{\prime}m}(a)
\displayfrac{\hat{a}_1^{\dag j+m^{\prime}}
\hat{a}_2^{\dag j-m^{\prime}}}
{\sqrt{(j+m^{\prime})!(j-m^{\prime})!}}\;,\nonumber\\ {\cal
U}(u)\displayfrac{\hat{a}_1^{j+m}\hat{a}_2^{j-m}}
{\sqrt{(j+m)!(j-m)!}} {\cal U}(u)^{-1} &=& e^{2i\alpha
j}\sum\limits_{m^{\prime}} D^{(j)}_{m^{\prime}m}(a)^*
\displayfrac{\hat{a}_1^{j+m^{\prime}}\hat{a}_2^{j-m^{\prime}}}
{\sqrt{(j+m^{\prime})!(j-m^{\prime})!}}\;,\nonumber\\ {\cal
U}(u)|j+m, j-m\rangle &=& e^{-2i\alpha j}\sum\limits_{m^\prime}
D^{(j)}_{m^{\prime}m}(a) |j+m^{\prime},
j-m^{\prime}\rangle\;,\nonumber\\
j&=&0,\;1/2,\;1,\ldots,\nonumber\\ m,\;m^{\prime}
&=&j,\;j-1,\;\ldots,\;-j\;;\nonumber\\ {\cal U}(u)
|\underline{z} \rangle &=& | u^*\;\underline{z}\rangle\;\cdot
\label{u2-action-ops}
\end{eqnarray}
\noindent
We have chosen the exponents of $\hat{a}$'s  and
$\hat{a}^{\dag}$'s, the numerical
factors, and the number operator eigenvalues, in such
a way that the results can be expressed neatly using the $SU(2)$
representation matrices in various unitary irreducible $SU(2)$
representations, namely the ${\cal D}$-functions of quantum
angular momentum theory~\cite{beidenharn-louck-I}.

To motivate the existence of several layers of classicality, we
now generalise the single mode $U(1)$-invariant real factorial
moments $\gamma_n$ of eqn.~(\ref{two-mode-defns-I}) to
two-mode quantities which
conserve total photon number and also transform in a closed and
covariant manner under $SU(2)$.  For this purpose, keeping in
mind eqns.~(\ref{u2-action-ops}),
it is convenient to start with the (in general
complex) classical monomials
 \begin{eqnarray}
f^j_{m_{1}m_{2}}\left(\underline{z}^{\dag},\underline{z}\right)
&=& N_{jm_{1}m_{2}}z_1^{*j+m_{1}}z_2^{*j-m_{1}}z_1^{j+m_{2}}
z_2^{j-m_{2}}\;,\nonumber\\ N_{jm_{1}m_{2}} &=&
[\left(j+m_1\right)!\left(j-m_1\right)!  \left(j+m_2\right)!
\left(j-m_2\right)!\;]^{-1/2}\;,\nonumber\\
j&=&0,\;1/2,\;1,\;\ldots,\nonumber\\ m_1,\;m_2&=&
j,\;j-1,\;\ldots,\;-j\;\cdot \end{eqnarray}
\noindent
The total power of $\underline{z}$ is equal to that of
$\underline{z}^{\dag}$, hence these are $U(1)$-invariant.  The
corresponding operators and their $SU(2)$ transformation laws
are:
 \begin{eqnarray}
f^j_{m_{1}m_{2}}\left(\underline{z}^{\dag},\underline{z}\right)
\rightarrow \widehat{F}^j_{m_{1}m_{2}} &=&N_{jm_{1}m_{2}}
\hat{a}_1^{\dag j+m_{1}}\hat{a}_2^{\dag j-m_{1}}
\hat{a}_1^{j+m_{2}}\hat{a}_2^{j-m_{2}}\;;\nonumber\\
a\;\epsilon\;SU(2) : {\cal U}(a)\widehat{F}^j_{m_{1}m_{2}} {\cal
U}(a)^{-1} &=& \sum\limits_{m^{\prime}_{1},m^{\prime}_{2}}
D^{(j)}_{m^{\prime}_{1}m_{1}}(a)
D^{(j)}_{m^{\prime}_{2}m_{2}}(a)^{*}\widehat{F}^j_{m^{\prime}_{1}
m^{\prime}_{2}}\;\cdot \end{eqnarray} For a given two-mode state
$\hat{\rho}$ we now generalise the factorial moments $\gamma_n$
of eqn.~(\ref{two-mode-defns-I}) to the following
three-index quantities:
 \begin{eqnarray}
\gamma^{(j)}_{m_{2}m_{1}}&=&\mbox{Tr}\left(\hat{\rho}\;\widehat{F}^j
_{m_{1}m_{2}}\right)\nonumber \\
&=&N_{jm_{1}m_{2}}\;\mbox{Tr}\left(\hat{\rho}\hat{a}_1^{\dag
j+m_{1}} \hat{a}_2^{\dag j-m_{1}}\hat{a}_1^{j+m_{2}}
\hat{a}_2^{j-m_{2}}\right)\cdot
\end{eqnarray}
\noindent
Their $SU(2)$ transformation law is clearly
\begin{eqnarray}
\hat{\rho}^{\prime} &=& {\cal U}(a)\;\hat{\rho}\;{\cal
U}(a)^{-1}\;: \nonumber\\ \gamma^{\prime (j)}_{m_{2}m_{1}} &=&
\sum_{m^{\prime}_{1},
m^{\prime}_{2}} D^{(j)}_{m^{\prime}_{1}m_{1}}\left(a^{-1}\right)
D^{(j)}_{m^{\prime}_{2}m_{2}}\left(a^{-1}\right)^*
\gamma^{(j)}_{m^{\prime}_{2}m_{1}^{\prime}}\;,\nonumber\\
\mbox{ie}\quad \gamma^{\prime (j)}&=& D^{(j)}(a)\;\gamma^{(j)}\;
D^{(j)}(a)^{\dag}\;\cdot \end{eqnarray}
\noindent
In the last line for each fixed $j$ the generalised moments
$\gamma^{(j)}_{m_{1}m_{2}}$ have been regarded as a (hermitian)
matrix of dimension $(2j+1)$.

On account of the fact that the total photon number is conserved
in the definition of these moments, calculation of
$\gamma^{(j)}_{m_{1} m_{2}}$ does not require complete knowledge of
$\phi(\underline{z})$ but only of a partly angle averaged
quantity ${\cal P}(I_1,I_2,\theta)$:
\begin{eqnarray}
\gamma^{(j)}_{m_{1}m_{2}}&=&N_{jm_{1}m_{2}}\int\limits^{\infty}_0
dI_1\;
\int\limits^{\infty}_0 dI_2\;\int\limits^{2\pi}_{0}
\displayfrac{d\theta}{2\pi}\nonumber \\
&&\quad {\cal P}(I_1,I_2,\theta)
(I_1I_2)^j(I_1/I_2)^{1/2(m_{2}+m_{1})}\cdot
e^{i(m_{1}-m_{2})\theta}\;,\nonumber \\ {\cal
P}(I_1,I_2,\theta)&=&\int\limits^{2\pi}_{0}
\displayfrac{d\theta_{1}}
{2\pi}\;\phi\left(I_1^{1/2}e^{i\theta_{1}},I_2^{1/2}
e^{i(\theta_{1}+\theta)}\right) \cdot \end{eqnarray}
\noindent
It is clear that these moments $\gamma^{(j)}_{m_{1}m_{2}}$ involve
more than just the photon number probabilities $p(n_{1},n_{2})$
which are just the ``diagonal'' case of the general matrix
element in eqn.~(\ref{examp-rho-matelements}):
 \begin{eqnarray}
p(n_{1},n_{2}) &=& \langle
n_1,n_2|\hat{\rho}|n_1,n_2\rangle\nonumber\\
&=&\int\limits^{\infty}_0 dI_1\;\int\limits^{\infty}_0
dI_2\;P(I_1,I_2)\;
e^{-I_{1}-I_{2}}I_1^{n_{1}}I_2^{n_{2}}\big/n_1!\;n_2!\;,\nonumber\\
P(I_{1},I_{2})
&=&\int\limits^{2\pi}_{0}\displayfrac{d\theta}{2\pi} \;{\cal
P}(I_1,I_2,\theta)\cdot \end{eqnarray}
\noindent
This is the two-mode version of eqn.~(\ref{prob-I}).  The subset of
``diagonal'' moments $\gamma^{(j)}_{mm}$ are calculable in terms of
$p(n_{1},n_{2})$ or $P(I_1,I_2)$:
\begin{eqnarray}
\gamma^{(j)}_{mm}&=&\int\limits^{\infty}_0
dI_1\;\int\limits^{\infty}_0 dI_2\;
P(I_1,I_2)I_1^{j+m}I_2^{j-m}\big/ (j+m)!\;(j-m)!\nonumber\\
&=&\sum\limits_{n_{1},n_{2}}\;p(n_{1},n_{2})\;n_1!\;n_2!\big/
(n_1-j-m)!(n_2-j+m)!(j+m)!(j-m)!  \end{eqnarray}

\noindent
However under a general $SU(2)$ mixing of the modes, the
expressions $\gamma^{(j)}_{mm},\;p(n_1,n_2),\;P(I_1,I_2)$ do not
transform in any neat way among themselves, and one is obliged
to enlarge the set to include the more general
$\gamma^{(j)}_{m_{1}m_{2}}$ and ${\cal P}(I_1,I_2,\theta)$. (In
particular, for these, the probabilities $p(n_{1},n_{2})$ are
inadequate).  When this is done we see the need to deal
with both the quantities
${\cal P}(I_1,I_2,\theta),\;P(I_1,I_2)$ derived
from $\phi(\underline{z})$ by a single or a double angular
average.  One can therefore distinguish four levels of
classicality for two-mode states:
\begin{eqnarray}
\hat{\rho}\;\mbox{classical}&\Leftrightarrow &
\phi(\underline{z})\geq 0,\;\left(\mbox{hence}\;
{\cal P}(I_1,I_2,\theta), P(I_1,I_2)\geq 0\right)\;;\nonumber\\
\hat{\rho}\;\mbox{semiclassical}\;I &\Leftrightarrow &
{\cal P}(I_1,I_2,\theta)\geq 0\;\left(\mbox{hence}\;
P(I_1,I_2)\geq 0\right),\;\mbox{but}\;
\phi(\underline{z})\not\geq 0\;;\nonumber\\
\hat{\rho}\;\mbox{semiclassical}\;II &\Leftrightarrow &
P(I_1,I_2)\geq 0\;\mbox{but}\;{\cal P} (I_1,I_2,\theta)\not\geq
0\;(\mbox{hence}\;\phi(z)\not\geq 0)\;; \nonumber\\
\hat{\rho}\;\mbox{strongly nonclassical} &\Leftrightarrow &
P(I_1,I_2)\not\geq 0 (\mbox{hence}\;\phi(\underline{z}),\; {\cal
P}(I_1,I_2,\theta)\not\geq 0)\cdot
\end{eqnarray}
These definitions can be cast in dual operator forms.  For example,
for semi classical -I states, we can say that for any classical
real nonnegative overall $U(1)$ phase invariant
$f\left(\underline{z}^{\dag},\underline{z}\right)$ the
corresponding operator $\widehat{F}$ has a nonnegative
expectation value, while this fails for some
$f\left(\underline{z}^{\dag},\underline{z}\right)$ outside this
class.  In the semi-classical-II case, we have to further limit
$f\left(\underline{z}^{\dag},\underline{z}\right)$ to be real
nonnegative and invariant under independent $U(1)\times U(1)$
phase transformations in the two modes, to be sure that the
expectation value of $\widehat{F}$ is nonnegative.

At this point we can see that these levels of classicality
possess different covariance groups. Since under a general
$U(2)\/$ transformation ${\cal U}(u), u\in U(2)\/$, the function
$\phi(\underline{z}) \/$ undergoes a point transformation,
$\phi(\underline{z}) \rightarrow
\phi^{\prime}(\underline{z}) = \phi(u^T \underline{z})\/$,
we see that the property of being classical is preserved by all
$U(2)\/$ transformations. On the other hand, a general $U(2)\/$
transformation can cause transitions among the other three
levels. The point transformation property is obtained for ${\cal
P}(I_1,I_2,\theta)\/$ and $P(I_1,I_2)\/$ only under the diagonal
$U(1) \times U(1)\/$ subgroup of $U(2)\/$; in fact
$P(I_1,I_2)\/$ is invariant under $U(1) \times U(1)\/$, while
${\cal P}(I_1,I_2,\theta)\/$ suffers a shift in the angle
argument $\theta$. Thus one can see that each of the three
properties of being semiclassical-I, semiclassical-II or
strongly non-classical is only $U(1) \times U(1) $ invariant.

As the number of modes increases further, clearly the hierarchy of
levels of classicality also increases.

Generalising inequalities of the form~(\ref{two-mode-defns-I})
for the diagonal quantities $\gamma^{(j)}_{mm}$, for
$\hat{\rho}$ classical or semiclassical-I or semiclassical-II,
is quite straightforward, since then we deal with the two modes
separately. The more interesting, and quite nontrivial, problem
is to look for matrix generalisations of
eqn.~(\ref{two-mode-defns-I}), bringing in the entire matrices
$\gamma^{(j)}=\left(\gamma^{(j)}_{m_{1}m_{2}}\right)$, and
looking for inequalities valid for states of the classical or
semi-classical-I types. (Of course for any quantum state
$\hat{\rho}$ we have the obvious property that $\gamma^{(j)}$,
for each $j$, is hermitian positive semidefinite.  This is the
two-mode generalisation of $\gamma_n\geq 0$ in the one-mode
case). However this is expected to involve use of the
Racah-Wigner calculus for coupling of tensor operators, familiar
from quantum angular momentum theory, inequalities for reduced
matrix elements, etc~\cite{biedenharn-louck-II}.

In the next Section we undertake a study of the particular case
$j=1$ which involves at most quartic expressions in $\hat{a}$'s
and $\hat{a}^{\dag}$'s.  This is just what is involved in giving
a complete account of the two-mode generalisation of the Mandel
Q-parameter familiar in the single-mode case.

\setcounter{equation}{0}
\section{Generalised photon-number fluctuation matrix
for two-mode fields}

For the one-mode case, with the single photon number operator
$\widehat{N}=\hat{a}^{\dag}\hat{a}$, we have some obvious
inequalities valid in all quantum states, and others valid in
semiclassical and classical states as defined in
eqn.~(\ref{cl-semicl-noncl}):

\begin{mathletters}
\begin{eqnarray}
\mbox{\underline{Any state}}:&&\nonumber \\
\label{gen-twomode-ineq-a}
\langle\widehat{N}\rangle &\equiv
&\mbox{Tr}\;(\hat{\rho}\widehat{N}) \equiv \gamma_1\geq 0\;;\\
\label{gen-twomode-ineq-b}
\langle \;:\widehat{N}^2\;:\rangle &\equiv & \mbox{Tr}\;\left(\hat{\rho}
\hat{a}^{\dag 2}\hat{a}^2\right)\equiv \gamma_2\geq 0\;;\\
\label{gen-twomode-ineq-c}
\langle\widehat{N}^2\rangle &\equiv & \mbox{Tr}\;
\left(\hat{\rho}\hat{a}^{\dag}\hat{a}\hat{a}^{\dag}\hat{a}\right)
\equiv \gamma_2 +\gamma_1 \geq 0\;;\\
\label{gen-twomode-ineq-d}
(\Delta N)^2&\equiv &\langle\widehat{N}^2\rangle
-\langle\widehat{N} \rangle^2\equiv
\langle(\widehat{N}-\langle\widehat{N}\rangle)^2
\rangle \equiv \gamma_2+\gamma_1-\gamma_1^2\geq 0\;;\\
\mbox{\underline{semiclassical or classical state}}\nonumber\\
\label{gen-twomode-ineq-e}
\langle :\;\widehat{N}^{2}\;:\rangle
-\langle\widehat{N}\rangle^2 &\equiv &(\Delta N)^2
-\langle\widehat{N}\rangle \equiv \gamma_2-\gamma_1^2\geq
0\;\cdot
\end{eqnarray}
\end{mathletters}
\noindent
(Here the dots : : denote normal ordering). The Mandel
Q-parameter is defined as~\cite{mandel-Q}
 \begin{equation}
Q\equiv \displayfrac{(\Delta N)^2-\langle\widehat{N}\rangle}
{\langle\widehat{N}\rangle}\equiv
\displayfrac{\gamma_2-\gamma_1^2}{\gamma_1}\;,
\end{equation}
\noindent
and it has the property of being nonnegative in classical
and semiclassical states. Conversely if $Q\/$ is negative,
the state is definitely strongly nonclassical.
The two cases $Q > 0\/$ and $Q < 0 \/$ correspond
respectively to super and subpoissonian photon number
distributions.

The inequalities (4.1) are not all independent, as some imply
others.  We now give the generalisation of these in matrix form,
to two-mode states.

We have to deal with four independent number-like operators
$\widehat{N}_{\mu}, \mu=0,1,2,3$ which we define thus:
 \begin{eqnarray} \widehat{N}_{\mu} &=&
\underline{\hat{a}}^{\dag}\sigma_{\mu}
\underline{\hat{a}} =
(\sigma_{\mu})_{rs}\hat{a}_r^{\dag}\hat{a}_s\;,
\nonumber\\ \hat{a}^{\dag}_r\hat{a}_s &=&
\frac{1}{2}(\sigma_{\mu})_{rs}\widehat{N}_{\mu}\cdot
\end{eqnarray}
\noindent
(Here $\sigma_0$ and $\sigma_j$ are the unit and the Pauli
matrices, and the sum on $\mu$ goes from 0 to 3).  The
expectation values of $\widehat{N}_{\mu}$ in a general state
$\hat{\rho}$ are written as $n_{\mu}\,$:
\begin{eqnarray}
\langle\widehat{N}_{\mu}\rangle \equiv \mbox{Tr}(\hat{\rho}
\widehat{N}_{\mu}) &=&\int d{\mu}(\underline{z})\phi(\underline{z})
\underline{z}^{\dag}\sigma_{\mu}\underline{z} =
n_{\mu}\;,\nonumber\\ \langle\hat{a}^{\dag}_r \hat{a}_s\rangle
&\equiv &
\mbox{Tr}\;\left(\hat{\rho}\hat{a}^{\dag}_r\hat{a}_s\right) =
\frac{1}{2}(\sigma_{\mu})_{rs}\;n_{\mu}.
\end{eqnarray}
\noindent
Thus $n_{\mu}$ and the matrix
$\gamma^{(1/2)}=\left(\gamma^{(1/2)}_{m_{1}m_{2}}\right)$
are essentially the same.  Since the $2 \times 2$ matrix
$\left(\langle\hat{a}^{\dag}_r\hat{a}_s\rangle\right)$ is
always hermitian positive semidefinite, we see that the
generalisation of inequality~(\ref{gen-twomode-ineq-a})
 to the two-mode case is

\begin{eqnarray}
n_0 - |\underline{n}|\geq 0\cdot
\label{vec-n-ineq}
\end{eqnarray}
\noindent
(All components of $n_{\mu}$ are real).
It may be helpful to remark that the matrix
$\gamma^{(1/2)}\/$ is analogous to the coherency matrix, and
the quantities $n_\mu \/$ to the Stokes parameters, in
polarisation optics~\cite{born-wolf}.

 Now we consider quadratic expressions in $\widehat{N}_{\mu}$
which are upto quartic in $\hat{a}^{\dag}_r$ and $\hat{a}_r$
combined.  To handle their normal ordering compactly, we first
define certain quadratic expressions in $\hat{a}_r$, and their
hermitian conjugates:
 \begin{eqnarray}
\widehat{A}_j &=& i\;\underline{\hat{a}}^T\sigma_2\;\sigma_j
\underline{\hat{a}}\;,\quad\widehat{A}^{\dag}_j =
-i\;\underline{\hat{a}}
^{\dag}\sigma_j\sigma_2\underline{\hat{a}}^*,\;
j=1,2,3\;;\nonumber\\
\hat{a}_r\hat{a}_s&=&-\frac{i}{2}\left(\sigma_j\sigma_2\right)_{rs}
\widehat{A}_j\;,\quad\hat{a}_r^{\dag}\hat{a}_s^{\dag}
=\frac{i}{2}(\sigma_2\sigma_j)_{rs}\;\widehat{A}^{\dag}_j\cdot
\end{eqnarray}
\noindent
Under the action of the unitary operators ${\cal U}(a)$
representing $SU(2)$, both $\widehat{A}_j$ and
$\widehat{A}_j^{\dag}$ transform as real three-dimensional
Cartesian vectors. Now we can easily express the result of
writing the product $\widehat{N}_{\mu}\widehat{N}_{\nu}$ as a
leading normally ordered quartic term plus a remainder:
 \begin{eqnarray}
\widehat{N}_{\mu}\widehat{N}_{\nu} &=&:\;\widehat{N}_{\mu}
\widehat{N}_{\nu}\;:\;+\left(\ell_{\mu\nu\lambda}+i\;
\epsilon_{0\mu\nu\lambda}\right)\;\widehat{N}_{\lambda}\;,\nonumber\\
:\;\widehat{N}_{\mu}\widehat{N}_{\nu}\;:&=& t_{\mu\nu
jk}\widehat{A} ^{\dag}_j\widehat{A}_k\;,\nonumber\\ t_{\mu\nu
jk} &=&\frac{1}{2} \left(\delta_{\mu\nu}\delta_{jk}-\delta_{\mu
j}\delta_{\nu k} -\delta_{\nu j}\delta_{\mu k} -i\;\delta_{\mu 0}
\epsilon_{0\nu j k}-i\;\delta_{\nu 0}\epsilon_{0\mu jk}\right)\;,\nonumber\\
\ell_{\mu\nu\lambda} &=& \delta_{\mu\nu}\delta_{\lambda 0}
+\delta_{\mu 0} \delta_{\nu\lambda} +\delta_{\nu
0}\delta_{\mu\lambda} -2\;\delta_{\mu 0}\;\delta_{\nu
0}\;\delta_{\lambda 0}\cdot \end{eqnarray}
\noindent
Here $\epsilon_{\sigma\mu\nu\lambda}$ is the four-index
Levi-Civita symbol with $\epsilon_{0123}=1$.  So the
anti-commutators and commutators among $\widehat{N}_{\mu}$ and
$\widehat{N}_{\nu}$ are:
\begin{mathletters}
\begin{eqnarray}
\frac{1}{2}\,\{\widehat{N}_{\mu},\widehat{N}_{\nu}\} &=& t_{\mu\nu j k}
\widehat{A}^{\dag}_j \widehat{A}_k +\ell_{\mu\nu\lambda}
\widehat{N}_{\lambda}\;,\\
\mbox{}[ \widehat{N}_{\mu},\widehat{N}_{\nu}]&=&
2\;i\;\epsilon_{0\mu\nu\lambda}\widehat{N}_{\lambda}
\end{eqnarray}
\end{mathletters}
\noindent
(These latter are just the $U(2)$ Lie algebra relations).  To
accompany $n_{\mu}$, in a general state we denote the
expectation values of $\widehat{A}^{\dag}_j \widehat{A}_k$ by
$q_{jk}$:
 \begin{eqnarray}
\langle\widehat{A}^{\dag}_j\widehat{A}_k\rangle \equiv
\mbox{Tr}\left(\hat{\rho}\widehat{A}_j^{\dagger}
\widehat{A}_{k}\right)=q_{jk}
\cdot
\label{q-jk}
\end{eqnarray}
\noindent
Clearly, $(q_{jk})$ is basically the matrix
$\gamma^{(1)}=(\gamma^{(1)}_{m_{1}m_{2}})$ and is always a $3
\times 3$ hermitian positive semidefinite matrix.  This
statement is the generalisation of
inequality~(\ref{gen-twomode-ineq-b}). We can also generalise
the
inequalities~(\ref{gen-twomode-ineq-c})~(\ref{gen-twomode-ineq-d})
by saying that for any quantum state the two matrices with
elements given by
\begin{mathletters}
\begin{eqnarray}
\langle\frac{1}{2}\{\widehat{N}_{\mu},\widehat{N}_{\nu}\}\rangle
&=& t_{\mu\nu jk} q_{jk} +\ell_{\mu\nu \lambda}n_{\lambda}\;,\\
\Delta(\widehat{N}_{\mu},\widehat{N}_{\nu})&\equiv & \frac{1}{2}
\langle \{ \widehat{N}_{\mu}-\langle\widehat{N}_{\mu}\rangle \;,\;
\widehat{N}_{\nu}
-\langle\widehat{N}_{\nu}\rangle \}\rangle\nonumber\\ &=&
t_{\mu\nu jk} q_{jk} + \ell_{\mu\nu\lambda}n_{\lambda}
-n_{\mu}\;n_{\nu}\, \end{eqnarray}
\end{mathletters}
\noindent
are both $4 \times 4$ real symmetric positive semidefinite.  As
in the one-mode case the  inequality obeyed by
$\left(\Delta \left( \hat{N}_{\mu}, \hat{N}_{\nu} \right) \right)\/$
implies the one obeyed by the anticommutator matrix
$\left( \langle \frac{1}{2}\left\{ \hat{N}_{\mu},\hat{N}_{\nu}
 \right\}\rangle\right) \/$

Now we search for matrix inequalities which are valid in
two-mode classical or semi-classical-I states, but not
necessarily in semi-classical-II or strongly nonclassical
states.  The key ingredient is the formula
\begin{eqnarray}
\langle\;:\;\frac{1}{2}\{\widehat{N}_{\mu},\widehat{N}_{\nu}\}\;:\;
\rangle &-& \langle \widehat{N}_{\mu}\rangle
\langle \widehat{N}_{\nu}\rangle \nonumber\\ &=&\int\
d\mu(\underline{z})\phi(\underline{z})
\left(\underline{z}^{\dag}\sigma_{\mu}\underline{z}-n_{\mu}\right)
\left(\underline{z}^{\dag}\sigma_{\nu}\underline{z}
-n_{\nu}\right) \nonumber\\
&=&\int\limits^{\infty}_{0}\int\limits^{\infty}_{0}
dI_1\;dI_2\int\limits^{2\pi}_{0}\displayfrac{d\theta}{2\pi}
{\cal
P}(I_1,I_2,\theta)\left(\underline{\zeta}^{\dag}\sigma_{\mu}
\underline{\zeta}-n_{\mu}\right)\left(\underline{\zeta}^{\dag}
\sigma_{\nu}\underline{\zeta}-n_{\nu}\right)\;,\nonumber\\
\underline{\zeta} &=&\left(\begin{array}{cc}I_1^{1/2} &\\
I_2^{1/2}&e^{i\theta}\end{array}\right)\cdot \end{eqnarray}
\noindent
We can now draw the following conclusion:

\underline{Classical or Semi-classical-I state}
 \begin{eqnarray} \left( \langle \;:
\;\frac{1}{2}\{\widehat{N}_{\mu},
\widehat{N}_{\nu}\}\;:\; \rangle \right.& - & \left. \langle
\widehat{N}_{\mu} \rangle
\langle \widehat{N}_{\nu}\rangle^{} \right) \nonumber\\ \equiv
\left(\Delta(\widehat{N}_{\mu},\widehat{N}_{\nu})
\right. &-& \left.
l_{\mu\nu\lambda}\langle\widehat{N}_{\lambda}\rangle
\right)\;\geq 0\cdot
\label{the-inequality}
\end{eqnarray}
\noindent
This is the intrinsic two-mode expression of super-poissonian
statistics, and its violation (possible only in
semi-classical-II or strongly nonclassical states) is an intrinsic
signature of two-mode subpoissonian photon statistics.
What makes this criterion nontrivial is the fact that for
any $n_\mu \/$ obeying eq.~(\ref{vec-n-ineq}) the $4\times4 \/$ matrix
$\left( l_{\mu\nu\lambda} n_\lambda \right)\/$ is real symmetric
positive semi-definite.

It is interesting to pin down the way in which this matrix
inequality~(\ref{the-inequality}) can go beyond a
single-mode condition~\cite{q-par-paper}.
The most general normalised linear
combination of the two mode-operators $\hat{a}_r$ is
determined by a complex two-component unit vector
$\underline{\alpha}$:
 \begin{eqnarray} \hat{a}(\underline{\alpha}) =
\underline{\alpha}^{\dag}
\underline{\hat{a}} &=& \alpha_r^*\hat{a}_r\;,\nonumber\\
\underline{\alpha}^{\dag}\underline{\alpha} &=& 1\;;\nonumber\\
\mbox{}[\hat{a}(\underline{\alpha})\;,
\hat{a}(\underline{\alpha})^{\dag}]&=& 1\cdot
\label{u2-alpha}
\end{eqnarray}
\noindent
For every such choice of a single mode, the inequality
{}~(\ref{the-inequality}) does imply the single-mode
inequality~(\ref{gen-twomode-ineq-e}).  We can see this
quite simply as follows.  Given $\underline{\alpha}$, we
define the real four-component quantity
$\xi_{\mu}(\underline{\alpha})$ by
\begin{eqnarray}
\xi_{\mu}(\underline{\alpha}) &=& \frac{1}{2}
\underline{\alpha}^{\dag}\sigma_{\mu}\underline{\alpha}\;:\nonumber\\
\xi_0(\underline{\alpha}) &=&
|\underline{\xi}(\underline{\alpha}) |=1/2\cdot \end{eqnarray}
\noindent
Then, using the completeness of $\sigma_{\mu}$ expressed by
 \begin{eqnarray} (\sigma_{\mu})_{rs}
\;(\sigma_{\mu})_{tu} = 2\;\delta_{ru}\;\delta_{st}\;,
\end{eqnarray}
\noindent
we have the consequences:
 \begin{eqnarray}
\xi_{\mu}(\underline{\alpha})\widehat{N}_{\mu}&=& \hat{a}
(\underline{\alpha})^{\dag}\hat{a}(\underline{\alpha})\equiv
\widehat{N}(\underline{\alpha})\;,\nonumber\\
\ell_{\mu\nu\lambda}\xi_{\mu}(\underline{\alpha})\xi_{\nu}
(\underline{\alpha})&=& \xi_{\lambda}(\underline{\alpha})\cdot
\end{eqnarray}
\noindent
Indeed we easily verify that (leaving aside $\xi_{\mu}=0$
identically)
 \begin{eqnarray}
l_{\mu\nu\lambda}\xi_{\mu}\xi_{\nu} =\xi_{\lambda}
&\Rightarrow&
\;\mbox{either}\;\;\xi_0=|\underline{\xi}|=1/2\nonumber\\
&\Leftrightarrow& \xi_{\mu} =
\frac{1}{2}\underline{\alpha}^{\dag}
\sigma_{\mu}\underline{\alpha}\;,\nonumber\\
&& \mbox{some}
\quad \underline{\alpha} \quad \mbox{obeying}\quad
\underline{\alpha}^{\dag}\underline{\alpha}=1\,,\nonumber\\
&\mbox{or}&\quad \xi_0=1\;,\;\underline{\xi}=0\cdot
\end{eqnarray}
\noindent
Saturating the left hand side of~(\ref{the-inequality})
with the latter
possibility, $\xi_{\mu}=\delta_{\mu 0}$, leads to the
superpoissonian condition for the total photon number
distribution.  Saturating it with
$\xi_{\mu}(\underline{\alpha})\xi_{\nu}(\underline{\alpha})$
we get as a consequence:
\begin{equation}
(\Delta \widehat{N}(\underline{\alpha}))^2 - \langle\widehat{N}
(\underline{\alpha}) \rangle \geq
0,\;\mbox{any}\;\underline{\alpha}\cdot
\label{variance}
\end{equation}
\noindent
In this way the two-mode matrix ``superpoissonian'' condition
{}~(\ref{the-inequality}) implies the scalar single mode
super poissonian condition
(4.1e) for every choice of normalised single mode with
annihilation operator $\hat{a}(\underline{\alpha})$, as well as
for the total photon number.

However, it is easy to see that {\it the information
contained in the matrix inequality~(\ref{the-inequality}) is not
exhausted by the collection of single mode
inequalities~(\ref{variance}) for all possible
choices of (normalized) $\underline{\alpha}$}.  Denoting
the real symmetric matrix appearing on the lefthand side
of ~(\ref{the-inequality}) by $(A_{\mu\nu})$,
 \begin{eqnarray}
A_{\mu\nu} = \Delta(\widehat{N}_{\mu},\widehat{N}_{\nu}) -
\ell_{\mu\nu\lambda}\langle \widehat{N}_{\lambda}\rangle\;,
\label{the-A-matrix}
\end{eqnarray}
\noindent
it is clear that
 \begin{eqnarray}
\xi_{\mu}\;A_{\mu\nu}\;\xi_{\nu} &\geq &0\;\mbox{for all}\;\;
\xi_{\mu}
\;\;\mbox{obeying}\;\xi_0=|\underline{\xi}|=1/2\nonumber\\
&\not \Rightarrow &(A_{\mu\nu})\geq 0
\label{eq-one-mode}
\end{eqnarray}
\noindent
Indeed, the lefthand side here reads in detail:
 \begin{eqnarray} \xi_{\mu} A_{\mu\nu}\xi_{\nu}
=\frac{1}{4}A_{00} + A_{0j}\xi_j + \xi_j \xi_k A_{jk}\;;
\end{eqnarray}
\noindent
and the nonnegativity of this expression for all 3-vectors
$\xi_j$ with $|\underline{\xi}|=1/2$ cannot exclude the
possibility of the $3 \times 3$ matrix $(A_{jk})$ having some
negative eigenvalues.  Part of the information contained in the
matrix condition~(\ref{the-inequality}) is thus irreducibly
two-mode in
character, a sample of this being:
\begin{eqnarray}
(A_{\mu\nu})\geq 0 \Rightarrow (A_{jk})\geq 0\cdot
\end{eqnarray}

Admittedly to a limited extent, this situation is analogous
to some well known properties of Wigner distributions.
Thus the marginal distributions in a single variable
obtained by integrating $W(q,p)\/$ with respect to $p\/$ or
with respect to $q\/$ (or any real linear combination of $q\/$ and
$p\/$) are always nonnegative probability distributions,
even though $W(q,p)\/$ is in general indefinite. So also
here, it can well happen that for a certain state both
$A_{00}\/$ and $\xi_{\mu}(\underline{\alpha}) A_{\mu \nu}
\xi_{\nu}(\underline{\alpha})\/$ are nonnegative for all
$\underline{\alpha}\/$ , yet $(A_{\mu \nu})\/$ is indefinite.

There exists in the literature a well known inequality for
two-mode fields, which when violated is a sign of
nonclassicality~\cite{lee-ineq}. It reads:
\begin{eqnarray}
\langle \hat{n}_1(\hat{n}_1-1) + \hat{n}_2(\hat{n}_2-1)-2
\hat{n}_1 \hat{n}_2 \rangle \geq 0, \nonumber \\
\hat{n}_1=\hat{a}_1^{\dagger}\hat{a}_1, \quad
 \hat{n}_2=\hat{a}_2^{\dagger}\hat{a}_2,
\label{lee-ineq}
\end{eqnarray}
and evidently involves only diagonal elements of the
matrix $(A_{\mu \nu})\/$. After rearranging the operators
in normal ordered form one can see that
\begin{eqnarray}
&&\langle \hat{n}_1(\hat{n}_1-1) + \hat{n}_2(\hat{n}_2-1)-2
\hat{n}_1 \hat{n}_2 \rangle \nonumber \\
&&=\langle \hat{a}_1^{\dagger 2}\hat{a}_{1}^{2}
+\hat{a}_2^{\dagger 2}\hat{a}_{2}^{2}-
2\/\hat{a}_1^{\dagger}\hat{a}_2^{\dagger}
\hat{a}_1 \hat{a}_2   \rangle \nonumber \\
&&=\frac{1}{2}(q_{11} + q_{22} -q_{33}) \nonumber \\
&&=A_{33} + n_3^{2} \nonumber \\
n_3&&= \langle \hat{a}_1^{\dagger}\hat{a}_1-
\hat{a}_2^{\dagger} \hat{a}_2  \rangle .
\label{lee-ineq-eq}
\end{eqnarray}
By our analysis, in any classical or semiclassical-I state
the matix $(A_{\mu \nu})\/$ is positive semidefinite, so
in particular $A_{33}\/$ and even more so the expression
$A_{33}+n_3^2 \/$, are both nonnegative. Thus the
inequality~(\ref{lee-ineq}) is certainly a necessary
condition for classical and semiclassical-I states.
Conversely, if~(\ref{lee-ineq}) is violated and
$A_{33}+n_3^2\/$ is negative, then certainly $A_{33}\/$ is
negative as well, and the state is either semiclassical-II
or strongly non-classical. However this condition is
unnecessarily strong since it asks for $A_{33}\/$ to be
less than $-n^2_{3}\/$; as we have shown, even the weaker
condition $A_{33} < 0\/$ is sufficient to imply that the
state is semiclassical-II or strongly nonclassical. Vice
versa, our necessary condition $A_{33} \geq 0\/$ for a classical
state or semiclassical-I state is stronger than the
condition~(\ref{lee-ineq}). In both directions, then, our
conditions are sharper than the ones existing in the
literature.

We conclude this Section by presenting a few examples
bringing out the content of the matrix
condition~(\ref{the-inequality}), in particular the
possibility of its containing more information than
all single-mode projections of it.
\begin{itemize}
\item[(a)] \underline{Pair-coherent states}: \\
These are  simultaneous
eigenstates of $\hat{a}_1 \hat{a}_2\/$ and
$\hat{a}_1^{\dagger}\hat{a}_1-
\hat{a}_2^{\dagger}\hat{a}_2\/$~\cite{pair-coh}:
\begin{eqnarray}
\hat{a}_1 \hat{a}_2 \vert \zeta,q\rangle &=& \zeta\vert
\zeta,q \rangle, \quad  \zeta \in \not C, \nonumber \\
(\hat{a}_1^{\dagger}\hat{a}_1-\hat{a}_2^{\dagger}\hat{a}_2)
\vert \zeta,q\rangle &=& q \vert \zeta,q\rangle , \quad
q=0, \pm1, \pm2 \cdots
\end{eqnarray}
For $q \geq 0\/$ these states are given by
\begin{equation}
\vert \zeta , q\rangle = N_q \sum_{n=0}^{\infty}
\frac{\zeta^{n}}{[n!(n+q)!]^{1/2 }}\vert n+q,n \rangle
\end{equation}
where $N_q\/$ is a normalisation constant.
It is known that in these states the second mode already
shows subpoissonian statistics~\cite{gsa-pair-coh}.
Thus if we write the
matrix~(\ref{the-A-matrix}) for these states as
$(A_{\mu \nu}(\zeta,q))\/$, then even
without having to nontrivially mix the modes we find:
\begin{eqnarray}
\underline{\alpha}=(0\,,\,1)^T, \quad
\xi_\mu(\underline{\alpha})&=&
\frac{1}{2}\,\underline{\alpha}^T \sigma_{\mu}\underline{\alpha}=
( \begin{array}{cccc} 1/2,&0,&0,&-1/2 \end{array}):
\nonumber \\
\xi_\mu(\underline{\alpha}) A_{\mu \nu}(\zeta,q)
\xi_\nu(\underline{\alpha}) &<& 0.
\end{eqnarray}
The matrix $A(\zeta,q)\/$ is indefinite and the pair
coherent states are therefore neither classical
nor even semiclassical-I. Consistent with this, a direct
numerical study of the least eigenvalue $
l\left(A(\zeta,q)\right)\/$ of$(A_{\mu
\nu}(\zeta,q))\/$ for sample values of $\zeta \/$ and
$q\/$, does show it to be negative.
\item[(b)] \underline{Two-mode squeezed vacuum}:
It has been shown elsewhere~\cite{two-mode-paper} that a two
mode squeezing transformation is characterised by two
independent intrinsic squeeze parameters $a\/$ and $b\/$ obeying
$a\geq b \geq 0 \/$. A representative of such a
transformation is
\begin{equation}
{\cal U}^{(0)}(a,b) =
\exp{\left[ \frac{(a-b)}{4}({\hat{a}_1^{\dagger 2}-
\hat{a}_1^{2}}) \right]}
\exp{\left[ \frac{(a+b)}{4}({\hat{a}_2^{\dagger 2}-
\hat{a}_2^{2}}) \right]}.
\end{equation}
The case $a=b\/$ essentially corresponds to the second mode
alone being squeezed. For general $a \neq b\/$ we have
genuine two-mode squeezing; while the
(Caves-Shumaker) limit $b=0\/$ involves
maximal entanglement of the two modes. We restrict our
analysis to this limit in the sequel. Then the two-mode
squeezed vacuum is characterised by the single parameter
$a\/$ and is
\begin{equation}
{\cal U}^{(0)}(a,0)\,\vert 0,0 \rangle =\exp{\left[
\frac{a}{4}\left( \hat{a}^{\dagger 2}_1 - \hat{a}^{2}_1+
\hat{a}^{\dagger 2}_2 - \hat{a}^{2}_2 \right)\right] \vert 0,0 \rangle
 }
\label{sq-vac}
 \end{equation}
The matrix $\left(A_{\mu \nu}(a)\right)\/$ can be explicitly
computed and happens to be diagonal:
\begin{eqnarray}
\left(A_{\mu \nu}(a)\right)=
\mbox{Diag}\left(\frac{1}{2}
\left( -3 + 7\,\cosh (2\,a) \right) \,{{{\rm Sinh}(a)}^2}
\,,\, 2\,\cosh (2\,a)\,{{{\rm Sinh}(a)}^2} \,,\right.
\nonumber \\  \left.
  -2\,{{{\rm Sinh}(a)}^2} \,,\,
   2\,\cosh (2\,a)\,{{{\rm Sinh}(a)}^2} \right)
\end{eqnarray}
We see that for all $a > 0\/$ this is indefinite, since
the third eigenvalue $A_{22}(a)\/$ is strictly negative.
This is displayed in Figure~1a, for $a\/$ in the range $ 0< a
<1 \/$. Thus for all $a >0\/$ the state~(\ref{sq-vac})
is definitely neither classical nor semiclassical-I. On the
other hand the leading diagonal element (eigenvalue)
$A_{00}(a)\/$ dominates the others in the sense that for
all choices of single mode the ``expectation value'' of
$A(a)\/$ is nonnegative:
\begin{equation}
\xi_\mu(\underline{\alpha}) A_{\mu \nu}(a)
\xi_\nu(\underline{\alpha}) \geq 0,\quad \mbox{all} \quad
\underline{\alpha} \nonumber \\
\end{equation}
Thus the squeezed vacuum~(\ref{sq-vac}) displays
nonclassicality via subpoissonian statistics in an
intrinsic or irreducible two-mode sense which never
shows up at the one mode level for any choice of that mode.
This is to be contrasted with the case of pair-coherent
states discussed previously. At the same time the
state~(\ref{sq-vac}) is also quadrature squeezed for all
$a>0\/$. Thus both these nonclassical features are present
simultaneously.
\item[(c)]\underline{Two-mode Squeezed Thermal state}:
This is defined as follows
(we again limit ourselves to the case $b=0\/$):
\begin{eqnarray}
\hat{\rho}(a,\beta)&=&{\cal
U}^{(0)}(a,0)\hat{\rho}_{0}(\beta) {\cal U}^{(0)}(a,0)^{-1},
\nonumber \\
\hat{\rho}_{0}(\beta)&=& (1-e^{-\beta})^2 \exp{\left[
-\beta(\hat{a}_1^{\dagger}\hat{a}_1 + \hat{a}_2^{\dagger}
\hat{a}_2) \right]},
\end{eqnarray}
At zero temperature $\beta \rightarrow \infty \/$ this
goes over to the previous
example (b).
Once again the matrix $\left(A_{\mu \nu}(a,\beta) \right)\/$
can be computed analytically and it turns out to be
diagonal:
\begin{eqnarray}
&&\left(A_{\mu \nu}(a, \beta) \right)=
{{\left( -1 + {e^{{\beta}}} \right) }^2} \times \nonumber \\
&&\mbox{Diag}\left(\begin{array}{c}
\frac{1}{8}\left(13 - 14\,{e^{{\beta}}} +
      13\,{e^{2\,{\beta}}} +
20\,(1 - {e^{2\,{\beta}}})\,\cosh (2\,a)
+ 7\,(1 +
{e^{{\beta}}})^2\,\cosh (4\,a)\right), \\
\frac{1}{2}\left((1 - {e^{{\beta}}})^2 + 2\,(1 -
     {e^{2\,{\beta}}})\,\cosh (2\,a) + (1 +
{e^{{\beta}}})^2\,\cosh (4\,a)\right)   , \\
 1 + {e^{2\,{\beta}}} + (1 -
      {e^{2\,{\beta}}})\,\cosh (2\,a), \\
  \frac{1}{2}\left((1 - {e^{{\beta}}})^2  + 2\,(1 -
      {e^{2\,{\beta}}})\,\cosh (2\,a) + (1 +
      {e^{{\beta}}})^2\,\cosh (4\,a)\right)
    \end{array} \right)
\end{eqnarray}
Now the third element $A_{22}(a,\beta)\/$ can become
negative for low enough temperature $T= \beta^{-1}\/$
or high enough squeeze parameter $a\/$. The variation of
the least eigenvalue $l\left(A(a, \beta)\right)\/$ of $A(a,
\beta)\/$ with respect to $a\/$ in the range $0 \leq a \leq
1\/$, for various choices of $\beta \/$, is shown in
Figures~(1b,c,d). One can see that if the temperature is not
too high, for sufficiently large $a\/$ the element
$A_{22}(a,\beta)\/$ becomes negative, indicating that the
state has then become semi-classical-II or strongly
nonclassical. (In comparison we recall that for quadrature
squeezing to set in the parameter $a\/$ must obey the
inequality $a >
\mbox{ln Coth\/}(\frac{\beta}{2})\/$~(\cite{two-mode-paper}) ) On the
other hand as in example (b), the leading element
$A_{00}(a,\beta)\/$ again dominates the others in the sense
that
\begin{eqnarray}
\xi_{\mu}(\underline{\alpha})\, A_{\mu \nu}(a,\beta) \,
\xi_{\nu}(\underline{\alpha}) \geq 0,\quad \mbox{all}\,
\underline{\alpha}
\end{eqnarray}
So once again, when $A_{22}(a, \beta)< 0\/$, the
subpoissonian statistics is irreducibly two-mode in
character. In Figures~(1b,c,d) we have also indicated the value
of the squeeze parameter $a\/$ at which quadrature squeezing
sets in. It is interesting to see that, for the states
described here, at each temperature, the irreducible two-mode
subpoissonian statistics occurs before squeezing. Therefore
( limiting ourselves to low order moments of
$\phi(\underline{z})$)
there exists a range of squeeze parameter where the only visible
nonclassicality is through such subpoissonian statistics.

The more general squeezed thermal state
\begin{equation}
\hat{\rho}(\beta,a,b)={\cal
U}^{(0)}(a,b)\hat{\rho}_{0}(\beta) {\cal U}^{(0)}(a,b)^{-1},
\end{equation}
has qualitatively similar properties. Detailed numerical
studies presented elsewhere~\cite{q-par-paper} have shown
that these states also do not show subpoissonian statistics
at the one-mode level. On the other hand, direct search for the
least eigenvalue of $\left(A_{\mu \nu}(a,b,\beta) \right)\/$
reveals that, for suitable values of $\beta, a,b$, this is
negative.

We thus have several instructive examples of the situation
indicated by eq.~(\ref{eq-one-mode})
\end{itemize}
\section{Concluding Remarks}
We have presented a dual operator and expectation value based
approach to the problem of distinguishing classical from
non-classical states of  quantised radiation, and thus
brought out the significance of this classification in a new
physically interesting manner. As the number of independent
modes increases, this approach leads to finer and yet finer
levels of nonclassical behaviour, in a steady progression. This
has been followed up by a complete analysis of photon number
fluctuations for two-mode fields, and a comprehensive concept of
subpoissonian statistics for such fields going beyond what can
be handled by techniques developed at the one-mode level.

In a previous paper we have set up the formalism needed to
examine the possibility of two-mode fields showing subpoissonian
statistics at the one-mode level in an invariant manner, by
following the variation of the Mandel Q-parameter as one
continuously varies the combination of the two independent modes
into a single mode. One can see through the work of the present
paper that that  preparatory analysis is a necessary prerequisite
to be able to pinpoint the aspects of subpoissonian statistics
which are irreducibly two-mode in character. Examples (b) and (c)
at the end of Section IV bring out this aspect vividly.

The inequality~(\ref{lee-ineq}) has been strengthened by our
approach to a sharper criterion to distinguish various
situations:
\begin{eqnarray}
\mbox{Classical or Semiclassical-I}\/ \Rightarrow A_{33} \geq 0;
\nonumber \\
A_{33} < 0 \Rightarrow \mbox{Semiclassical-II or strongly
nonclassical}.
\label{lee-repeat}
\end{eqnarray}
{}From eqns.~(\ref{the-A-matrix},~\ref{lee-ineq-eq})
we see that $A_{33}\/$ has the following neat
expression:
\begin{eqnarray}
A_{33} &=& \left(\Delta \hat{n_1}\right)^2 -\langle\hat{n_1} \rangle
+\left(\Delta \hat{n_2}\right)^2 -\langle\hat{n_2} \rangle
-2 \Delta \left( \hat{n}_1,\hat{n}_2 \right) \nonumber \\
&=&\langle (\hat{n}_1-\hat{n}_2)^2 \rangle
-(\langle \hat{n}_1-\hat{n}_2\rangle)^2-
\langle \hat{n}_1+\hat{n}_2\rangle
\end{eqnarray}
It is thus expressible solely in terms of expectations and
fluctuations of the original (unmixed) mode number
operators $\hat{n}_1,\hat{n}_2\/$ and their functions.  One
can now see easily, again from
equation~(\ref{the-A-matrix}), that the
statements~(\ref{lee-repeat}) are part of a wider set of
statements involving only expectations of functions of
$\hat{n}_1,\hat{n}_2\/$:
\begin{mathletters}
\begin{eqnarray}
A_{00} &=& \left( \Delta \hat{N}_0 \right)^2 -
\langle \hat{N}_0 \rangle,
\nonumber \\
A_{03}&=&A_{30}=\Delta\left( \hat{N}_0,\hat{N}_3 \right) -\langle
\hat{N}_3 \rangle,
\nonumber \\
A_{33} &=&
\left(\Delta \hat{N}_3 \right)^2 -\langle \hat{N}_0 \rangle,
\nonumber \\
\hat{N}_0&=&\hat{n}_1+\hat{n}_2\quad, \quad
\hat{N}_3=\hat{n}_1-\hat{n}_2;
 \\
\mbox{Classical or semiclassical-I}\/ &\Rightarrow&
\left( \begin{array}{cc}
A_{00}& A_{03}\\
A_{30}& A_{33}
\end{array}\right) \geq 0;
\nonumber\\
\left( \begin{array}{cc}
A_{00}& A_{03}\\
A_{30}& A_{33}
\end{array}\right)<0 \quad &\Rightarrow& \mbox{Semiclassical-II or
strongly  nonclassical}.
\end{eqnarray}
\end{mathletters}
All other inequalities involving matrix elements such as
$A_{01}, A_{02},A_{13}\cdots\/$ involve ``phase sensitive''
quantities going beyond $\hat{n_1}\/$ and $\hat{n_2}\/$.

Going back to the matrix $A=\left( A_{\mu \nu} \right)\/$, we see
that from its properties we cannot immediately distinguish
between the classical and semiclassical-I situations, or between
the semiclassical-II and strongly non-classical situations. In both
the former, $A\/$ is positive semidefinite; while if $A\/$ is
indefinite, one of the latter two must occur. It would be
interesting, for pair coherent states or  squeezed thermal states
for instance, to be able to see, when $A\/$ is indefinite, whether
we have a semi-classical-II or a strongly non-classical state,
and whether this depends  on and varies with the parameters in the
state.

{\large \bf Acknowledgements}\\
Arvind thanks University Grants Commission, India for financial
support.


\begin{figure}
\caption{Plots of least eigenvalue of the matrix $(A_{\mu \nu})\/$
as function of squeeze parameter $a\/$. Figure1(a)
displays the least eigenvalue of $\left( A_{\mu \nu} \right)$
for squeezed vacuum where as Figures1(b, c, and d) display
the same for squeezed thermal states with inverse
temperature $\beta$ taking the values $4.0, 2.0\/$ and
$1.0\/$ respectively. In Figures1(b,c, and d) the arrow
shows the setting in of quadrature squeezing.}
\end{figure}
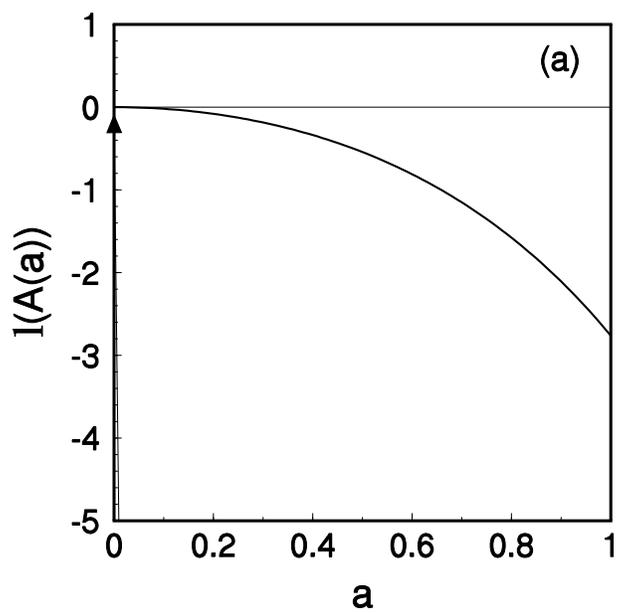
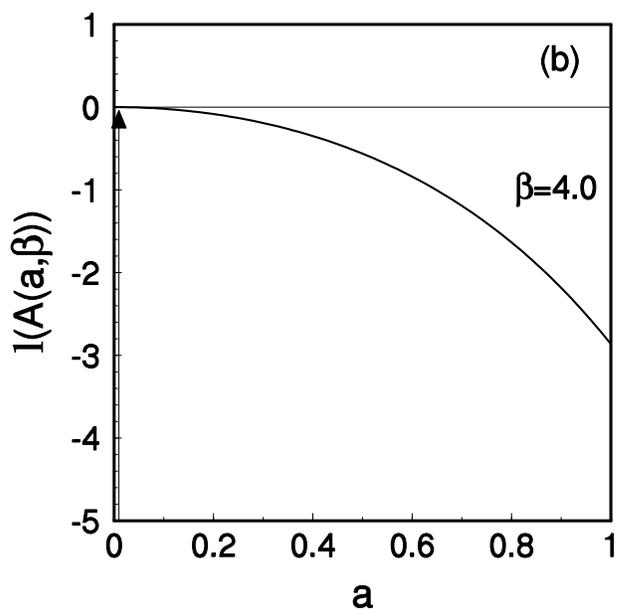
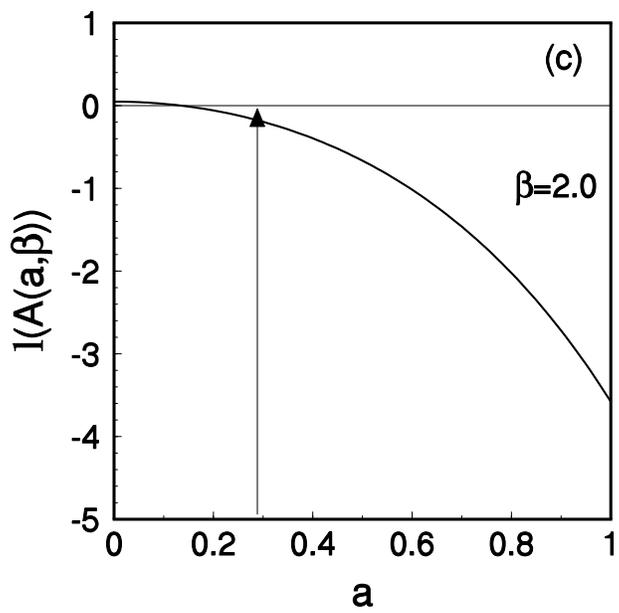
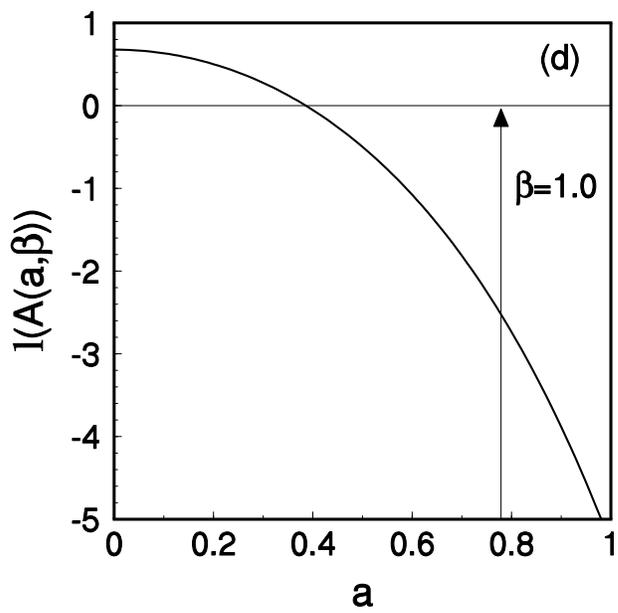
\end{document}